# Technical Note: A fast and monolithic prototype clinical proton radiography system optimized for pencil beam scanning

Ethan A. DeJongh[1], Don F. DeJongh[1*], Igor Polnyi[1], Victor Rykalin[1], Christina Sarosiek[2+], George Coutrakon[2], Kirk L. Duffin[3], Nicholas T. Karonis[3,4], Caesar E. Ordoñez[3], Mark Pankuch[5], John R. Winans[3], James S. Welsh[6,7]

[1]ProtonVDA LLC, Naperville, IL 60563, USA
[2]Department of Physics, Northern Illinois University, DeKalb, IL 60115, USA
[3]Department of Computer Science, Northern Illinois University, DeKalb, IL 60115, USA
[4]Argonne National Laboratory, Data Science and Learning Division, Argonne, IL 60439, USA
[5]Northwestern Medicine Chicago Proton Center, Warrenville, IL 60555, USA
[6]Edward Hines Jr VA Medical Center, Radiation Oncology Service, Hines, IL 60141, USA
[7]Department of Radiation Oncology, Loyola University Stritch School of Medicine, Maywood, IL 60153, USA

*Senior Author

+Corresponding Author: csarosiek1@niu.edu, 1425 W Lincoln Hwy, DeKalb, IL 60115

**ABSTRACT**

Purpose: To demonstrate a proton imaging system based on well-established fast scintillator technology to achieve high performance with low cost and complexity, with the potential of a straightforward translation into clinical use.

Methods: The system tracks individual protons through one (X, Y) scintillating fiber tracker plane upstream and downstream of the object and into a 13 cm-thick scintillating block residual energy detector. The fibers in the tracker planes are multiplexed into silicon photomultipliers (SiPMs) to reduce the number of electronics channels. The light signal from the residual energy detector is collected by 16 photomultiplier tubes (PMTs). Only four signals from the PMTs are output from each event, which allows for fast signal readout. A robust calibration method of the PMT signal to residual energy has been developed to obtain accurate proton images. The development of patient-specific scan patterns using multiple input energies allows for an image to be produced with minimal excess dose delivered to the patient.

Results: The calibration of signals in the energy detector produces accurate residual range measurements limited by intrinsic range straggling. We measured the water-equivalent thickness (WET) of a block of solid water (physical thickness of 6.10 mm) with a proton radiograph. The mean WET from all pixels in the block was 6.13 cm (SD 0.02 cm). The use of patient-specific scan patterns using multiple input energies enables imaging with a compact range detector.

Conclusions: We have developed a prototype clinical proton radiography system for pretreatment imaging in proton radiation therapy. We have optimized the system for use with pencil beam scanning systems and have achieved a reduction of size and complexity compared to previous designs.

**INTRODUCTION:**

ProtonVDA LLC has developed a prototype clinical proton radiography system for pretreatment patient setup and range verification in proton radiation therapy. Kohler[1] first proposed proton radiography in 1968 as a high contrast imaging modality. Since then, researchers have shown that proton radiography can be used as a pretreatment quality assurance tool in proton therapy. In particular, it can be used as a proton range check to detect differences between the actual and the predicted proton range from anatomical changes[2] and, in addition, to align the patients to the treatment beam[3]. Recent studies have explored the potential of proton radiographs to reduce uncertainties due to patient-specific Hounsfield units to relative stopping powers conversion used with x-ray CT scans[4,5].

The detector system, shown in Figure 1, described in this technical note tracks single protons of known incident energy and direction from low-intensity pencil beams. Tracker planes measure the location of individual protons upstream and downstream of the object. The steering of the pencil beam scanning system determines the direction of the incoming proton. After traversing the object, the residual energy of the proton is measured in a scintillating block energy detector with a set of photomultiplier tubes (PMT).

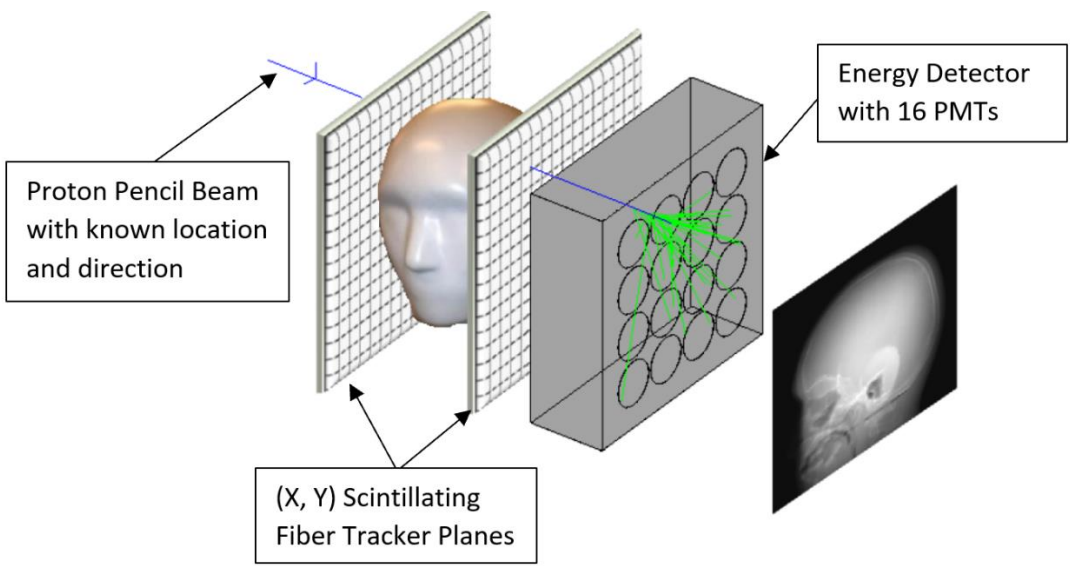

*Figure 1: Diagram depicting the radiography detector system. The proton source is on the left side of the image, and the pencil beam has a known location at isocenter. The proton passes through the upstream scintillating fiber tracker plane, the phantom, the downstream scintillating fiber tracker plane, then stops in the scintillating block of the energy detector. The scintillating block converts the residual energy of the proton into light, which is collected by 16 PMTs located on the downstream face of the scintillating block.*

The detector system is designed to minimize the size and complexity of the hardware such that it can easily be integrated into proton therapy treatment rooms. The use of pencil beam scanning allows for the use of a single upstream tracker plane and a reduction of the delivered dose delivered as described in Sections 2.1, 2.3, and 3.2. A calibration procedure to obtain WEPL from PMT signals has been developed, which allows for fast signal readout and minimal electronics, as described in Section 2.2.

**2. METHODS AND MATERIALS:**

**2.1 Tracker planes**

The proton radiography system includes one set of (X, Y) tracker planes in front of the object and one set after the object. The sensitive area of the tracker planes is 38.4 x 38.4 cm$^2$. Each X and Y tracker plane consists of two layers in beam's eye view of 1 mm$^2$ square scintillating fibers (Saint-Gobain, model BCF12S-S1.00N). The layers are offset by one half fiber width, as illustrated in Figure 2a. One fiber from each layer is connected into a fiber pair. Twelve bunches of 32 fiber pairs laid side-by-side creates one X or Y tracker plane. The number of electronics is reduced by feeding one fiber pair from each bunch into a single 6 x 6 mm$^2$ solid state silicon photomultiplier (SiPM) (Hamamatsu, model S13360-6050VE), as illustrated in Figure 2b. There are a total of 128 SiPMs for the entire proton radiography system (32 X and 32 Y for front tracker, 32 X and 32 Y for rear tracker). The signal flow from the scintillating fibers to the data acquisition (DAQ) box is shown in Figure 3. An accelerator plan defines the location of each pencil beam at isocenter and helps to determine which of the 12 fiber bunches were hit. The fiber signals then determine the actual location of the proton hit on the tracker plane at a 0.5 mm pitch. We then calculate the entrance angle of the proton on the upstream tracker from the location of the proton hit and the distance from the steering magnets in the accelerator.

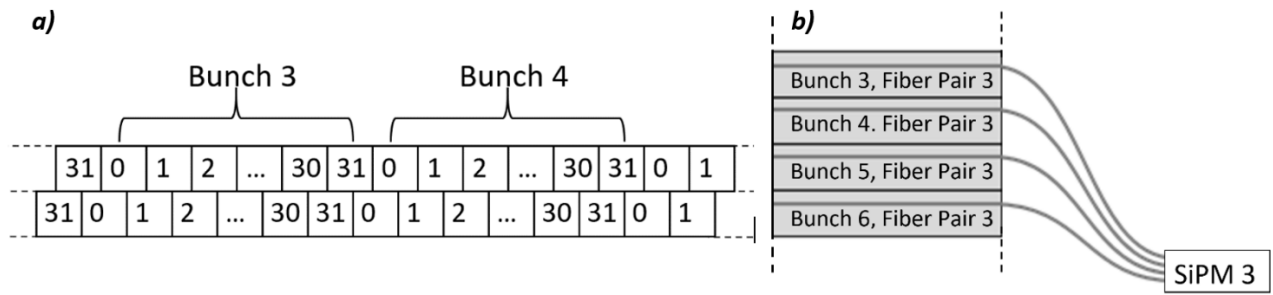

Figure 2: a) Cross-section of an X or Y tracker plane consisting of two layers in beam's eye view of scintillating fibers each 1 x 1 mm$^2$ by 40 cm active length. The fibers are grouped into 12 bunches of 32 fiber pairs where each pair includes one fiber from each layer. All fibers from every bunch with the same numerical label are attached to a single silicon photomultiplier (SiPM), which has an active area of 6 x 6 mm$^2$. b) Top view of four adjacent fiber bunches. One fiber pair from each bunch is read out through a common SiPM. So, each SiPM reads out 12 fiber pairs regularly spaced across the plane.

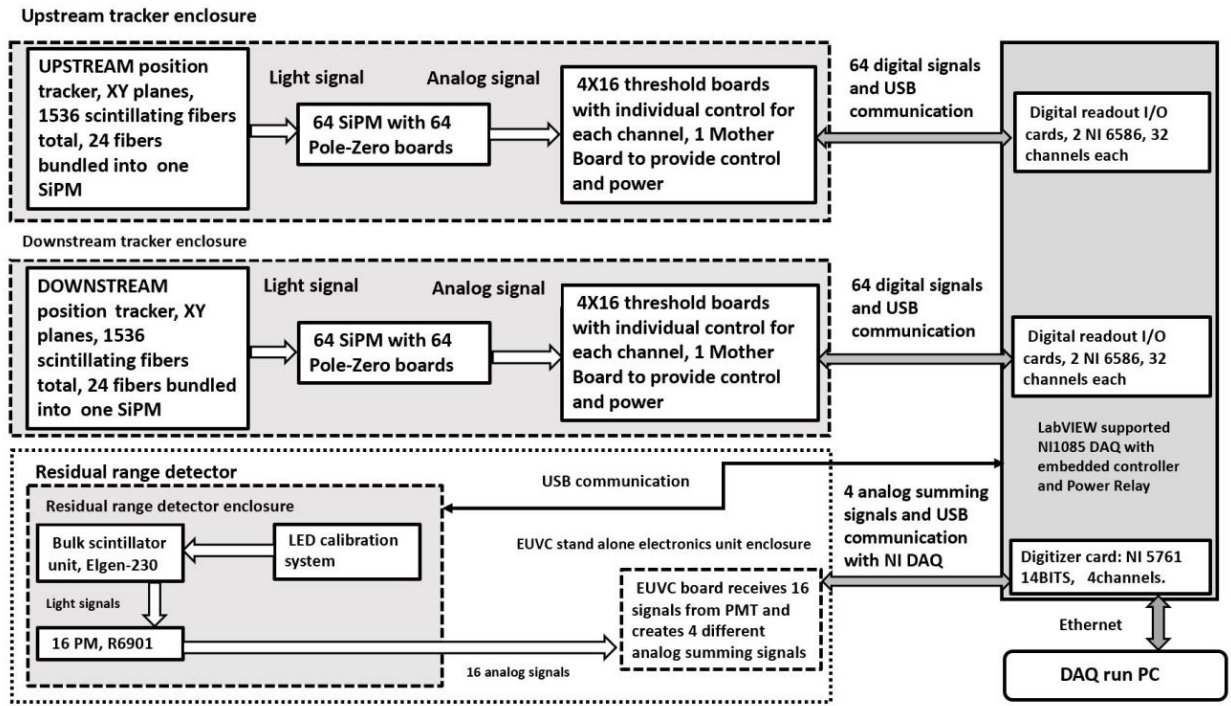

*Figure 3. Block diagram of the signal flow from the trackers and energy detector to the National Instruments (NI) DAQ box.*

Several other collaborations developed and built proton imaging tracking planes utilizing scintillating fibers coupled to photodetectors[6,7,8]. The designs described in Lo Presti[6] and Pemler[7] both couple multiple fibers to a single SiPM channel to reduce the number of electronics while the NIU[8] detector independently reads out triplet bundles of 0.5 mm fibers. Our design couples fibers from both layers into a fiber pair, then couples regularly spaced fiber pairs to a single SiPM, for 24 fibers coupled to one channel. Each X or Y plane requires 32 channels and a full tracker plane requires 64 channels, as shown in Figure 2. In addition, we use only two planes (one X and one Y) for each tracking plane, rather than the four planes (two X and two Y) in the similar designs. The major difference between the former designs and our design is in the pre-determination of the location of the proton hit from the known spot pattern. The Pembler and Presti detectors require detection in two channels to resolve ambiguities in the location of the proton hit, while our design only requires one detection and resolves the ambiguities with apriori knowledge of the intended position of the pencil beam.

**2.2 The energy detector and calibration to WEPL**

The energy detector contains a compact 40 cm x 40 cm-transverse size, 13 cm-thick scintillator block (Eljen Technology, model EJ230) and sixteen 76 mm diameter vacuum photomultiplier tubes (PMTs) (Hamamatsu, model R6091). The scintillating block has a sensitive thickness of 10 cm, and the radiographic image is acquired from several proton scans using different energies in different regions of the object. The use of several energies ensures that there will be some protons in each image pixel that pass entirely through the object and have a measurable residual water-equivalent range less than 10 cm. For example, the variations in WET in the pediatric head phantom (shown as the phantom in Figure 1) used in this project (Computerized Imaging Reference Systems, Inc., model HN-719) require a minimum of three energies for a complete anterior-posterior (AP) proton radiograph. A 4 x 4 grid of PMTs located on the downstream face of the scintillating block collect the scintillating photons. The PMTs cover about 58% of the surface of the energy detector. The scintillator sides not covered by the PMTs, i.e. the sides parallel to the central beam axis and the upstream face, are painted black to absorb photons, and the PMTs collect only direct photons that have not scattered off the walls. This approach minimizes the collection time of photons.

The residual range of each proton is measured and converted to WEPL using the following method. The 16 PMTs on the backend of the energy detector measure the light output generated by the protons that passed through the object. The light output depends on the residual energy and the position of the proton with respect to the energy detector. The position dependence arises from the spacing of the PMTs, where a proton that is directed at the center of a PMT will have a stronger signal than a proton directed between two PMTs, as demonstrated in Figure 4a. An electronics board combines the 16 PMT signals into four weighted-sum signals, named E, U, V, and C, minimizing the data volume output. The E signal is the total sum of all 16 PMT signals. For most positions, the E signal increases approximately linearly with the residual range of the proton but with a position-dependent slope, as shown in Figure 4b. The position of the proton entering the energy detector is known from the tracker planes. Therefore, the E signal alone gives an accurate measure of the residual range for most protons entering the detector. However, some positions have non-linear slopes in the E signal at large residual ranges, and many protons will undergo multiple Coulomb scattering throughout the energy detector, confounding the linear E-signal dependence. For these cases, the U, V, and C signals provide additional information to improve the residual range accuracy. The U and V signals are weighted sums of the PMT signals along diagonals of the 4 x 4 array, as shown in Figures 5a and 5b, respectively. The U and V signals are designed to give an estimate of position and help correct for multiple scattering inside the detector. Lastly, the C signal is a weighted sum of the PMT signals based on the two concentric squares in the 4 x 4 array, as shown in Figure 5c. The C signal helps determine the residual range of the proton in the areas where the U and/or V signals are near zero. For example, the C signal becomes more positive with depth for events near the center, where the U and V signals have low absolute values. The chosen weights gave the best residual range resolution out of several options that we tested in detector simulations. For example, increased weights for the U and V signals near the corners (-3, -2, -1, 0, 1, 2, 3 format) performed better than taking the difference between two sums (-1, -1, -1, 0, 1, 1, 1 format).

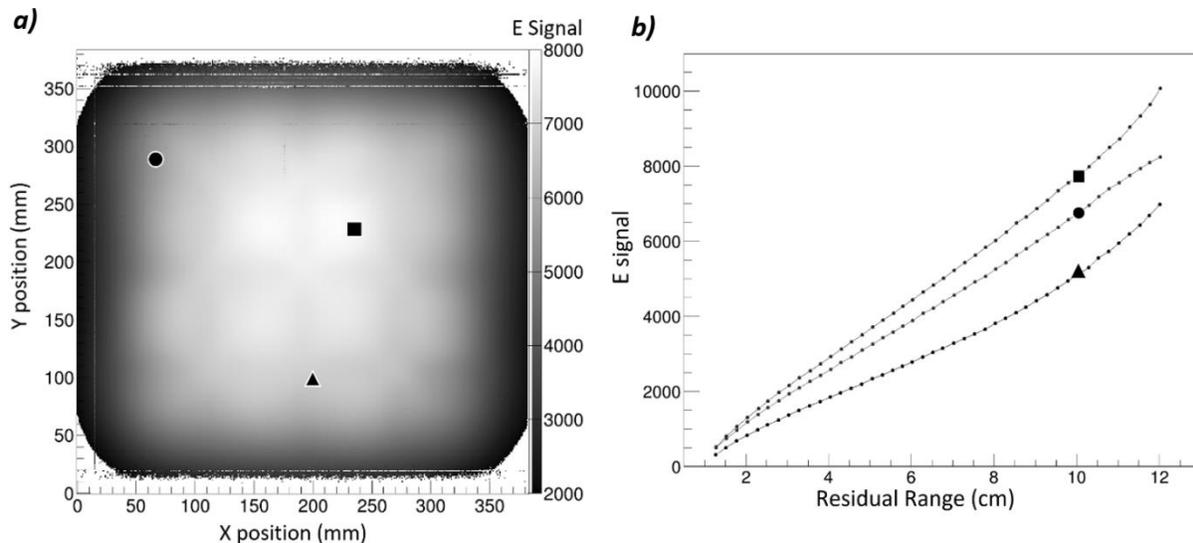

Figure 4: a) Spatial map of the E signal, i.e., the sum of the 16 PMT signals, for a uniform scan of protons that have a 10 cm residual range in the energy detector. The spacing of the PMTs create a position dependence in the E signal. Note that the noise at the boundaries come from being at the edge of the beam field and therefore less data. b) Plot of the E signal versus residual range for the three (X, Y)

positions indicated by different symbols in Figure 4a. The E signal increases with residual range because a higher energy particle entering the scintillator block will produce more light, and, therefore, a higher overall signal. For additional details, see text.

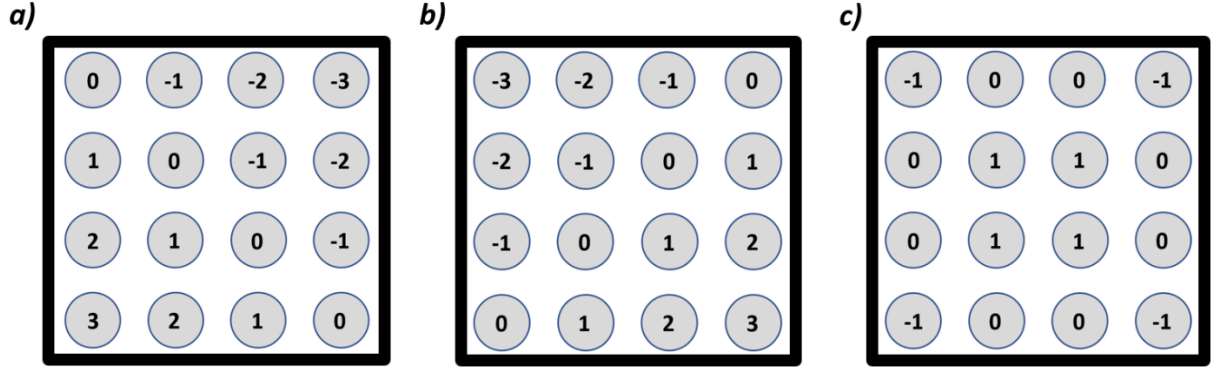

Figure 5: Weights of the individual PMT signals for a) the U signal, b) the V signal, and c) the C signal.

A calibration procedure was developed to convert the E, U, V, and C signals to residual proton range. Protons are delivered to a 30 x 40 cm² field at 44 different residual ranges between 1.25 and 12 cm. Approximately 10,000 protons were delivered to each spot spaced 5 mm in the transverse direction and 0.25 cm in depth. The proton data is binned into a 3D calibration grid with points for each of the 384 X and Y fiber positions and the 44 residual ranges from the proton data in the calibration scans. The E, U, V, and C signals for each proton are measured and the average E, U, V, and C signals for each point in the calibration grid (denoted as $\bar{E}, \bar{U}, \bar{V}, \bar{C}$, respectively) is calculated. From these data, the EUVC covariance matrix $K_{EUVC}$ is calculated and collected into the calibration grid of (X, Y, Residual Range). The covariance matrix for each point in the grid is defined as

$$K_{EUVC} = \begin{bmatrix} cov(EE) & cov(EU) & cov(VE) & cov(EC) \\ cov(UE) & cov(UU) & cov(UV) & cov(UC) \\ cov(VE) & cov(VU) & cov(VV) & cov(VC) \\ cov(CE) & cov(CU) & cov(CV) & cov(CC) \end{bmatrix},$$

where the covariance between two variables $A$ and $B$ is $cov(AB) = \left(\frac{1}{N}\right)\sum_{i=1}^{N}(A_i - \bar{A})(B_i - \bar{B})$ and $N$ is the number of protons corresponding to a grid point[9,10].

We now use the values in the calibration grid to determine the residual range of each proton in the imaging data set. Therefore, for a single proton in the image data set, we calculate the χ² versus residual range for all points in the calibration grid having the same (X, Y) position. We start with the χ² definition:

$$\chi^2 = \Delta^T K_{EUVC}^{-1} \Delta \quad (4)$$

and

$$\Delta = \begin{bmatrix} E - \bar{E} \\ U - \bar{U} \\ V - \bar{V} \\ C - \bar{C} \end{bmatrix} \quad (3)$$

(2)

where the averages and covariance matrix are taken from the calibration data and E, U, V, C are the measurements from a single proton in the image data. The three lowest χ² values versus range are fitted with a parabolic function, and the residual range of the proton is the location of the minimum of the fit. The proton WEPL is calculated as the difference between the water-equivalent range of the incident proton and the residual range of the proton measured by the energy detector.

The image reconstruction software bins the WEPL values into pixels at the isocenter plane, which is the plane normal to the beam located at a defined distance from the accelerator scanning magnets. The water equivalent thickness (WET) is calculated as the most likely WEPL value in each pixel and displayed as a grey-scale proton radiographic image. The image reconstruction methods used to produce a proton radiographic image from the WEPL data produced by this system are described in detail by Ordoñez, et. al.[11]

### 2.3 Patient-specific scan pattern

For proton radiography, the useful data comes only from protons that pass entirely through the object and stop in the energy detector. The Bragg peak of the protons occurs in the energy detector, and the dose delivered to the patient is part of the entrance region of the proton depth-dose curve. Due to the compact nature of the residual energy detector, energy modulation is required to ensure protons pass through the object and have a residual water-equivalent range of less than 10 cm at every (X, Y) location. In clinical practice, dose-minimization is achieved by developing a patient-specific scan pattern. A previously acquired x-ray CT image of the patient can be used to determine the appropriate beam energy for a specific (X, Y) position. The Hounsfield units are converted to proton relative stopping power, and a digitally reconstructed WET image is created. This image is then used to determine the set of energies for the proton radiography scan as a function of (X, Y) positions. The goal of this method is to maximize the number of imaging protons that stop in the energy detector and minimize the Bragg peaks stopping inside of the patient. In clinical use, this will minimize the imaging dose to the patient.

## 3. RESULTS

### 3.1 Calibration Accuracy

A proton radiograph of a block of solid water with known WET of 6.10 cm was taken to determine the accuracy of the calibration procedure. About 12 million protons with an energy of 128 MeV were uniformly scanned across a 30 x 40 $cm^2$ field size. The resultant radiograph is shown in Figure 6a. The WEPL from each proton in the image is displayed in a histogram in Figure 6b. The mean WEPL from the protons used in the radiograph is 6.12 cm (SD 0.26 cm), as shown in the histogram in Figure 6b. Each pixel included approximately 100 protons, and the mean WET from all pixels in the radiograph is 6.13 cm (SD 0.02 cm), as shown in the histogram in Figure 6c.

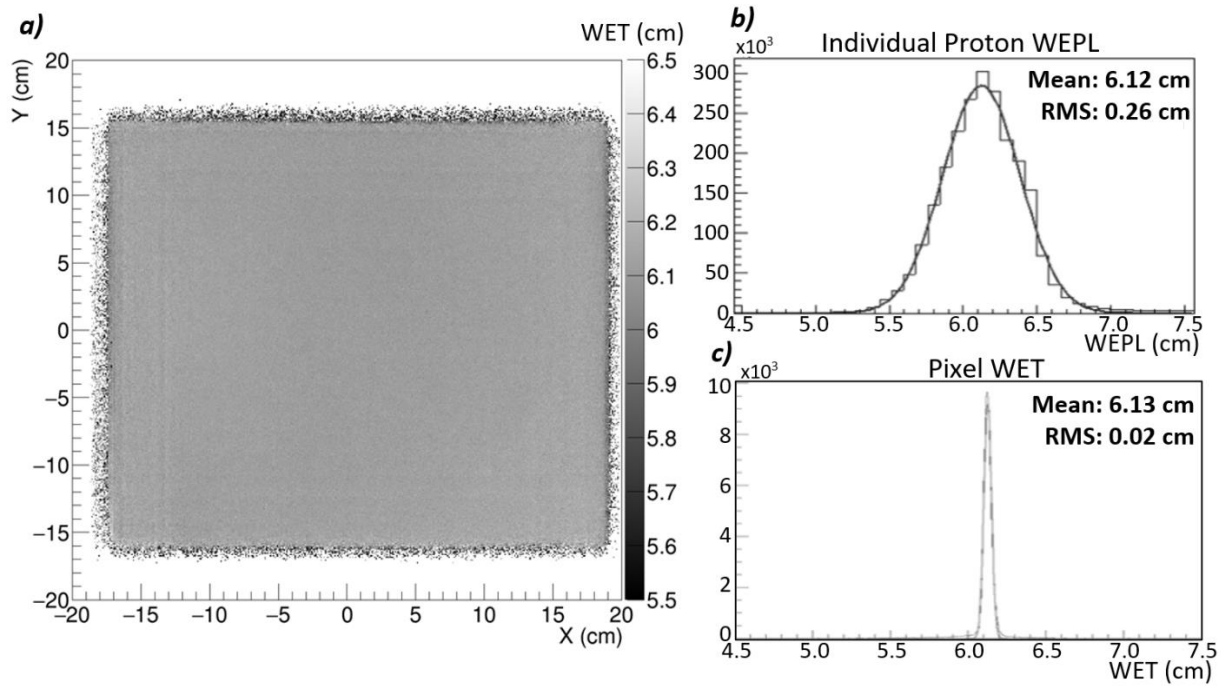

*Figure 6: a) Proton radiograph of a 6.10 cm thick block of solid water using 128 MeV protons. b) Histogram with the best-fit normal distribution of the measured WEPL values of individual protons resulting in a mean value of 6.12 mm (SD 0.26 mm). c) Histogram with the best-fit normal distribution of the reconstructed average WET values per pixel resulting in a mean value of 6.13 mm (SD 0.02 mm).*

### 3.2 Imaging Dose

A simulation was performed with TOPAS[12] version 3.1 to compare the dose deposited in the pediatric head phantom with a full-field uniform scan spot pattern and a patient-specific scan pattern. The simulation included three energies (180 MeV, 140 MeV, and 100 MeV) and a total imaging field size of 20 cm x 20 cm. Each pencil beam spot included 2,500 simulated protons, and the spots were spaced 0.5 cm apart, resulting in about 100 protons per 1 mm$^2$ pixel. The dose deposited in a selected axial slice of the phantom using the two different scan techniques is compared in Figure 7. Figure 7a shows the dose deposition from the full-field uniform scan. Many of the protons from the lower two energies stop in the phantom. The data from these protons are not used in the image reconstruction and, therefore, are wasted dose to the patient or phantom. By using patient-specific scan patterns, such as in Figure 7b, the dose is reduced by a factor of eight in some regions as seen in Figure 7c, and the majority of the dose delivered is deposited by protons that are used in the image reconstruction. The resultant radiograph from the simulated patient-specific scan pattern is shown in Figure 8. The reconstruction algorithm, described in Ordoñez, et. al., identifies the appropriate energy (energies) for each pixel and eliminates protons with other energies[11]. As long as the patient-specific scan has adequate coverage, the resulting radiograph will be identical to those created with full-field uniform scans.

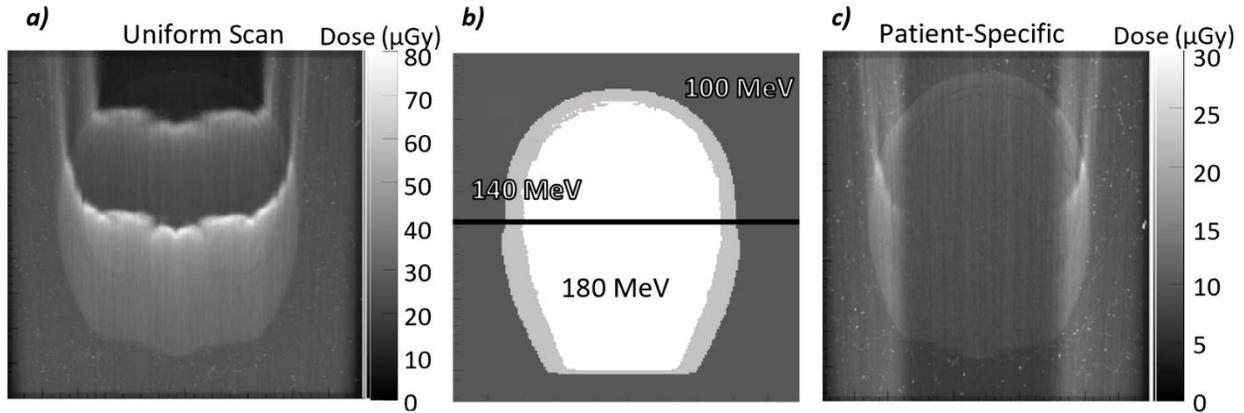

*Figure 7: Dose distribution in a single axial slice of the pediatric head phantom from a proton radiograph with three energy scans. The proton beam is coming from the bottom of the image. a) The phantom is scanned for each energy separately, and many protons in the lower energy scans stop in the phantom. The 100 MeV protons have the shortest range and deposit greater dose midway through the phantom. The 140 MeV protons deposit greater dose in the thickest portion of the phantom. The 180 MeV protons pass completely through the phantom and deposit very little dose in the phantom. However, in the regions outside the phantom will have too high of residual energy to stop in the energy detector. b) Diagram depicting the locations of the 100 MeV pencil beams (dark grey), 140 MeV pencil beams (light grey), and 180 MeV pencil beams(white) in the patient-specific scan pattern for an anterior-posterior (AP) proton radiograph. At these locations, most of the protons will pass completely through the phantom, minimizing the deposited dose. c) Dose distribution from the patient-specific scan pattern shown in Figure 7b. Note that different grey scales were used for Figures 7a and 7c.*

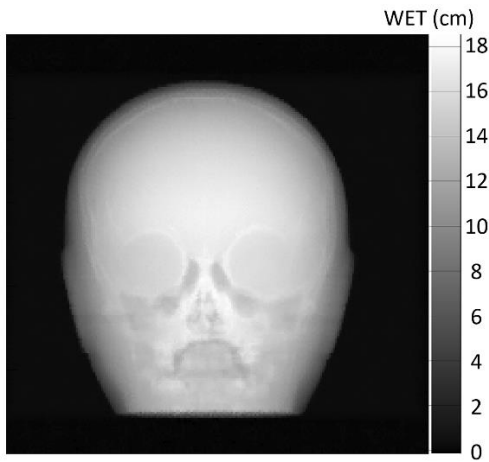

*Figure 8: Simulated proton radiograph using the patient-specific scan pattern shown in Figure 7b and Figure 7c. The radiograph was reconstructed using the image reconstruction techniques described in Ordoñez, et. al.[11]. The reconstruction algorithm automatically removes protons that did not stop in the energy detector. This means that the proton radiograph created with the full-field uniform scan will be identical.*

## 4. DISCUSSION

The proton radiography system described in this technical note has a minimal amount of electronics, is low-weight, and compact, making it straightforward to translate the system into clinical use. The system is designed to operate with proton intensities up to 10 MHz, a factor of five higher than other systems such as the proton CT detector described by Sadronzinski, et. al.[13]. A 20 x 20 cm$^2$ field size with 100 protons per 1 mm$^2$ pixel can be delivered in less than one second. The system provides a fast and efficient method of imaging for pretreatment range verification and alignment procedures.

The use of only a single tracker plane upstream and downstream of the detector, as compared to two planes in initial designs, provides considerable volume savings; however, it removes the ability to measure incident and exit angles used during image reconstruction. With the use of scanned pencil beams, the incident angle can be calculated from the knowledge of the source-to-isocenter distance and the actual position of the pencil beam in the isocenter plane. There is, however, a reduction in spatial resolution due to multiple Coulomb scattering in the object, which randomizes the exit angle of the protons. A direct measurement of the proton angle would reduce the uncertainty of the path of the proton and increase spatial resolution. On the other hand, detection of range errors, may not require such a high degree of spatial resolution, justifying this cost-saving design.

The use of a compact monolithic residual energy detector further reduces the weight and size of the system. By using pencil beams and patient-specific scan patterns, the dose delivered to the patient is minimized to micro-Gy levels. Adjusting the energy according to object thickness reduces the nuclear interactions, which are eliminated during the data preprocessing step, thus reducing unnecessary imaging dose to the patient. On the other hand, the use of lower-energy protons will result in more extensive multiple Coulomb scattering and a further reduction of spatial resolution. As described previously, the use of an image reconstruction method for proton radiography that employs a most likely path algorithm where the exit angle is unknown can partially compensate for this loss of spatial resolution[11,14]. Simulations comparing the use of one downstream tracker plane to two downstream tracker planes indicate the uncertainty of the path estimation is typically 0.3 mm larger for the single plane case. Future work will study the impact of this error on the final image.

## 5. CONCLUSION

We have described a proton radiography detector that can be translated into clinical use. It has been developed to be implemented on beamlines using low-intensity pencil beam scanning such that the same proton delivery system is used for pretreatment imaging, patient setup verification and subsequent treatment. A calibration procedure was developed for fast conversion of energy signals to water equivalent pathlength with a minimum amount of electronics. The use of the pencil beam scanning system for imaging allows for further reduction of the energy detector while allowing for optimizing the imaging dose.

**ACKNOWLEDGEMENTS**

This work used resources at Northwestern Medicine Chicago Proton Center. The authors thank Dr. Reinhard Schulte at Loma Linda University for reviewing and commenting on this paper. We also thank Nick Detrich from Ion Beam Applications for creating control software to deliver proton spot patterns at the correct intensity and energies required by the radiography system.

This work was sponsored by the National Cancer Institute of the National Institutes of Health contract numbers R44CA203499 and R44CA243939.

**CONFLICT OF INTEREST STATEMENT**

The authors have intellectual property rights to the innovations described in this paper. James S. Welsh has served as a medical advisor to ProTom International. Don F. DeJongh and Victor Rykalin are co-owners of ProtonVDA LLC.